\documentclass[a4paper,11pt]{article}
\usepackage{style/pos}
\usepackage[separate-uncertainty=true, range-units=single]{siunitx}
\usepackage{xspace}
\usepackage{amssymb}
\usepackage{amsmath}
\usepackage{sidecap}

\newcommand{\eV}[1]{$10^{#1}\,$eV\xspace}
\newcommand{\Egeo}{\ensuremath{E_\mathrm{geo}}\xspace}
\newcommand{\dmax}{\ensuremath{d_\mathrm{max}}\xspace}
\newcommand{\rhomax}{\ensuremath{\rho_\mathrm{max}}\xspace}
\newcommand{\ace}{\ensuremath{a_\mathrm{ce}}\xspace}
\newcommand{\fgeo}{\ensuremath{f_\mathrm{geo}}\xspace}
\newcommand{\fgeopos}{\ensuremath{f_\mathrm{geo}^\mathrm{pos}}\xspace}
\newcommand{\fce}{\ensuremath{f_\mathrm{ce}}\xspace}
\newcommand{\fgs}{\ensuremath{f_\mathrm{GS}}\xspace}

\title{Reconstructing inclined extensive air showers from radio measurements}
\ShortTitle{Radio reconstruction of inclined air showers}

\author*[a,b]{Tim Huege}
\author[a,c]{Felix Schlüter}

\affiliation[a]{Institute for Astroparticle Physics (IAP), Karlsruhe Institute of Technology, Karlsruhe, Germany}

\affiliation[b]{Astrophysical Institute, Vrije Universiteit Brussel, Brussels, Belgium}

\affiliation[c]{Universidad Nacional de San Martín, Instituto de Tecnologías en Detección y Astropartículas,\\ Buenos Aires, Argentina}

\emailAdd{tim.huege@kit.edu; felix.schlueter@kit.edu}

\abstract{We present a reconstruction algorithm for extensive air showers with zenith angles between 65$^\circ$ and 85$^\circ$ measured with radio antennas in the 30-80 MHz band. Our algorithm is based on a signal model derived from CoREAS simulations which explicitly takes into account the asymmetries introduced by the superposition of charge-excess and geomagnetic radiation as well as by early-late effects. We exploit correlations among fit parameters to reduce the dimensionality and thus ensure stability of the fit procedure. Our approach reaches a reconstruction efficiency near 100\% with an intrinsic resolution for the reconstruction of the electromagnetic energy of well below 5\%. It can be employed in upcoming large-scale radio detection arrays using the 30-80 MHz band, in particular the AugerPrime Radio detector of the Pierre Auger Observatory, and can likely be adapted to experiments such as GRAND operating at higher frequencies.}

\FullConference{37$^{\rm{th}}$ International Cosmic Ray Conference (ICRC 2021)\\
		July 12th -- 23rd, 2021\\
		Online -- Berlin, Germany}

\begin{document}
\maketitle

\section{Introduction}

Radio detection of inclined air showers with zenith angles beyond 65$^\circ$ has recently come into focus because it allows measurements of cosmic rays up to the highest energies \cite{HuegeUHECR2014}. As radio measurements provide pure information on the electromagnetic energy of air showers, they ideally complement particle detector measurements which at these angles measure the (almost) pure muon content of air showers. This led the Pierre Auger Collaboration to equip its complete Surface Detector array with radio antennas \cite{PontIcrc2019}. Another experiment that will focus on radio-based measurements of inclined air showers, induced by cosmic rays or neutrinos, will be GRAND \cite{GRANDWhitePaper}.

In this work, we present a signal model and reconstruction algorithm for air showers measured with radio antennas in the 30-80 MHz band, evolved from our previous work \cite{HuegeARENA2018,HuegeIcrc2019}. It will be directly applicable for the AugerPrime Radio Detector, and its methodology should be adaptable to detectors with different observing frequency ranges and ambient conditions such as GRAND.

\section{Core-shift, early-late correction and symmetrization}

\begin{figure}
    \centering
    \includegraphics[width=0.49\textwidth]{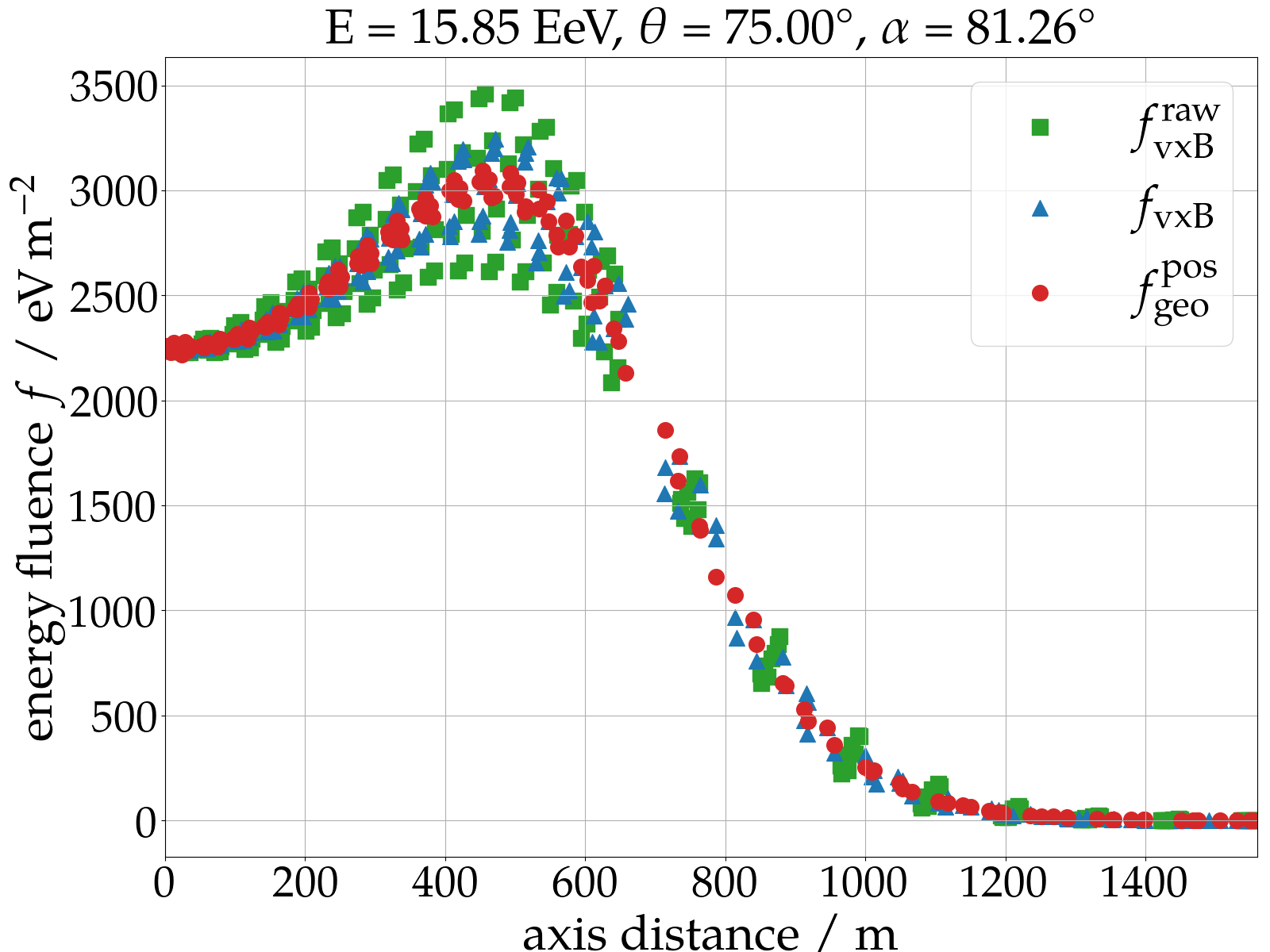}\hfill
    \includegraphics[width=0.49\textwidth]{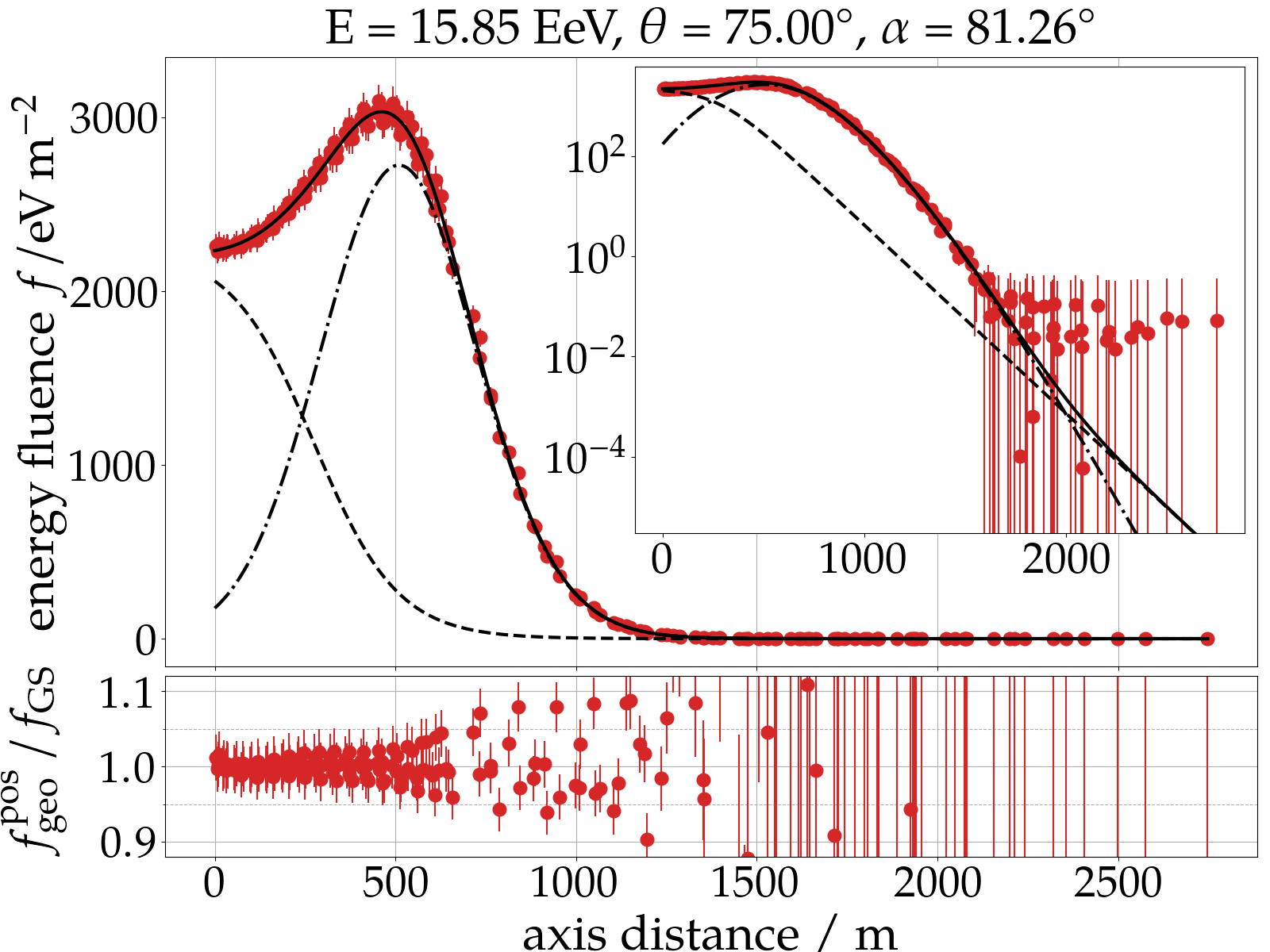}
    \caption{Lateral signal distribution of an example iron shower with a zenith angle of 75$^\circ$. Left: The raw energy fluence in the $\bf{v} \times \bf{B}$ polarization (green squares) is highly asymmetric even after transformation to the shower plane and accounting for a core shift by atmospheric refraction. Early-late correction improves the symmetry (blue triangles). Following decomposition of the total fluence according to eqn.\ (\ref{eqn:posfraction}), the geomagnetic energy fluence \fgeopos (red circles) then is rotationally symmetric. Right: Example \fgeopos distribution and its fit with eqn.\ (\ref{eqn:fgs}). The bottom panel shows the residuals. The contributions of the Gaussian and sigmoid components of the fit are shown as well. Data points at largest distances are affected by particle thinning.}
    \label{fig:LDFs}
\end{figure}

The footprint of the radio-emission energy fluence measurable on the ground is highly asymmetric. It can be symmetrized by a transformation to the shower plane, followed by a geometrical early-late correction (see \cite{HuegeIcrc2019} for details), and finally a calculation of the pure geomagnetic energy fluence \fgeo by removal of the fluence contributed by the charge-excess emission \fce. This decomposition is done by exploiting the known polarization characteristics of the two contributions, governed by the polar angle $\phi$ in the shower plane at every antenna position \cite{GlaserJCAP2016}:
\begin{gather}
\label{eq:ce_fluence_pos2}
    f_{\mathrm{ce}}^{\mathrm{pos}} = \frac{1}{\sin^2(\phi)} \cdot  f_{\textbf{v}\times\textbf{v}\times\textbf{B}}. \nonumber \\
    f_{\mathrm{geo}}^{\mathrm{pos}} = \left(\sqrt{f_{\textbf{v}\times\textbf{B}}} -  \frac{\cos(\phi)}{|\sin(\phi)|} \cdot \sqrt{f_{\textbf{v}\times\textbf{v}\times\textbf{B}}} \right)^2\label{eqn:posfraction}
\end{gather}
where ${f_{\textbf{v}\times\textbf{B}}}$ and $f_{\textbf{v}\times\textbf{v}\times\textbf{B}}$ denote the fluences in the given polarizations, $\textbf{v}$ and $\textbf{B}$ denoting the shower axis and magnetic field vector. The pure \fgeopos then has a rotationally symmetric ``lateral distribution'' as shown in Figure \ref{fig:LDFs} (left). Please note that compared with our previous work, we now allow for a systematic core shift with respect to the Monte Carlo core (and particle core) that occurs in inclined air showers due to refraction in the atmosphere \cite{SchlueterCoreshift2020} while applying these symmetrization steps.

\section{Rotationally symmetric LDF}

The rotationally symmetric geomagnetic fluence distribution \fgeo can then be fit with a one-dimensional ``lateral distribution function'' (LDF). In contrast to our earlier work we now use a two-component LDF. The first component is a Gaussian with a width $\sigma$ centered at an axis distance $r_0$ (the exponent $p$ is 2 for $r < r_0$ but decreases slowly for larger axis distances, see below; this was inspired by the model put forth in reference \cite{GlaserGeoCE}). The second component is a ``sigmoid'' that fills in the inner part of the LDF and has a relative amplitude $a_\mathrm{rel}$ with respect to the Gaussian component as well as shape parameters $s$ and $r_{02}$: 

\begin{equation}
        \fgs(r) = f_0 \left[\exp\left(-\left(\frac{r - r_0}{\sigma} \right) ^ {p(r)} \right)  + \frac{a_\mathrm{rel}}{1 + \exp\left(s \cdot \left(\frac{r}{r_0} - r_{02} \right)\right)} \right] \label{eqn:fgs}
\end{equation}

The achievable fit quality can be judged in Figure \ref{fig:LDFs} (right). At the largest axis distances, increased fluence values are present due to artifacts arising from particle thinning. We adopt a relative fluence uncertainty of 3\% for every antenna plus an absolute uncertainty of 0.01\% of the maximum geomagnetic fluence of the given simulation to ensure that the data points affected by thinning do not negatively influence the fit. Our new fit function performs visibly better than our previously chosen function, especially in the outer part of the LDF, cf.\ Figure 7 in reference \cite{HuegeIcrc2019}.

\section{Exploiting LDF parameter correlations with air shower characteristics}

\begin{figure}
    \centering
    \includegraphics[width=0.85\textwidth]{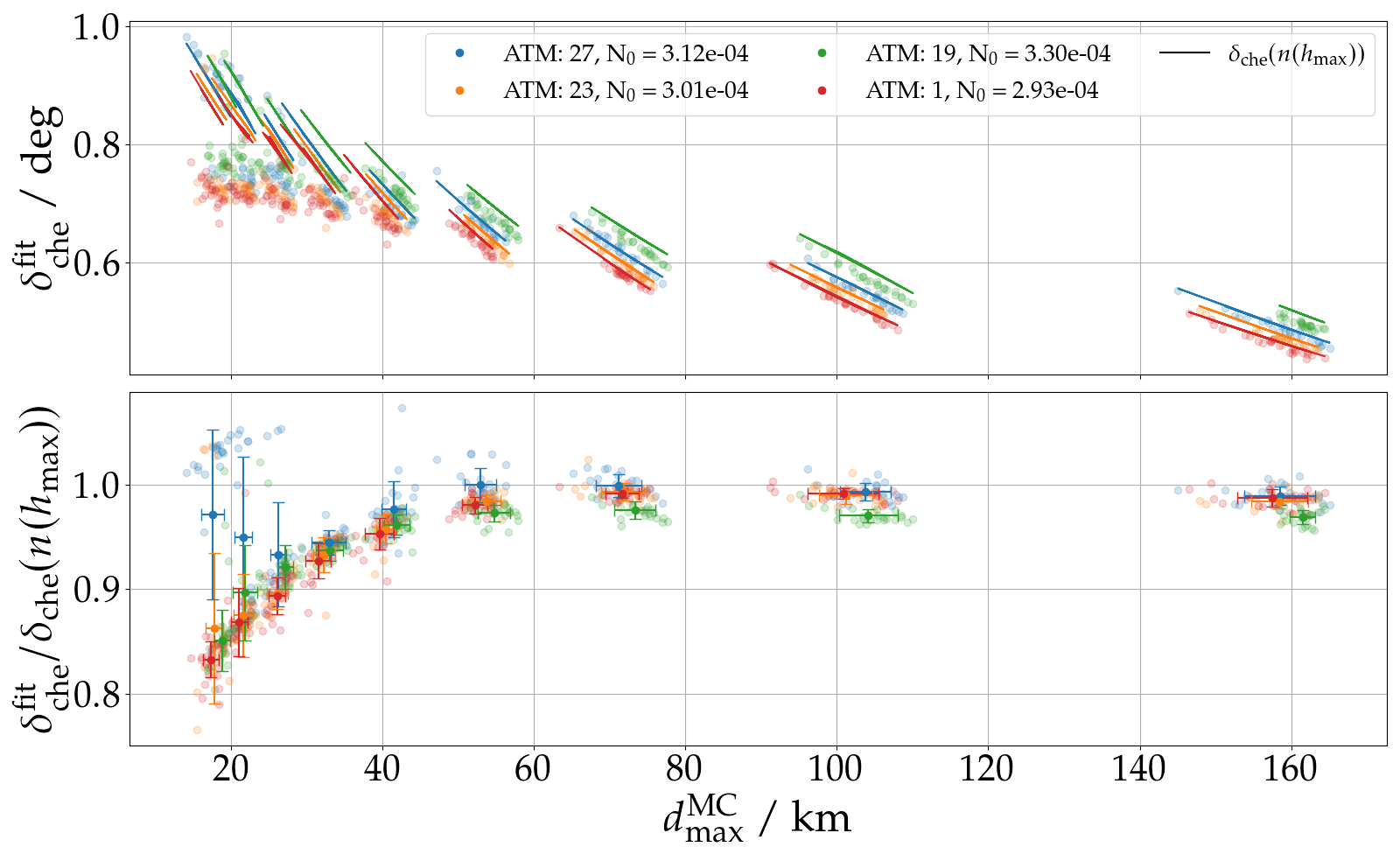}
    \caption{Top: Cherenkov angle $\delta_\mathrm{che}$ as calculated from the refractive index $n$ at shower maximum by eqn.\ (\ref{eqn:Cherenkov}) (lines) compared to the angle $\delta_\mathrm{che}$ calculated from $r_0$ extracted from fitting eqn.\ (\ref{eqn:fgs}) to individual showers (points). Simulations with four different atmospheric models are shown (pre-defined in CORSIKA 7, with refractivity at sea level set to the value quoted as $N_0$). Bottom: Deviation between fitted and calculated values. The profiles show the mean and standard deviations of the fitted values.}
    \label{fig:cherenkovangle}
\end{figure}

\begin{figure}
    \centering
    \includegraphics[width=0.245\textwidth]{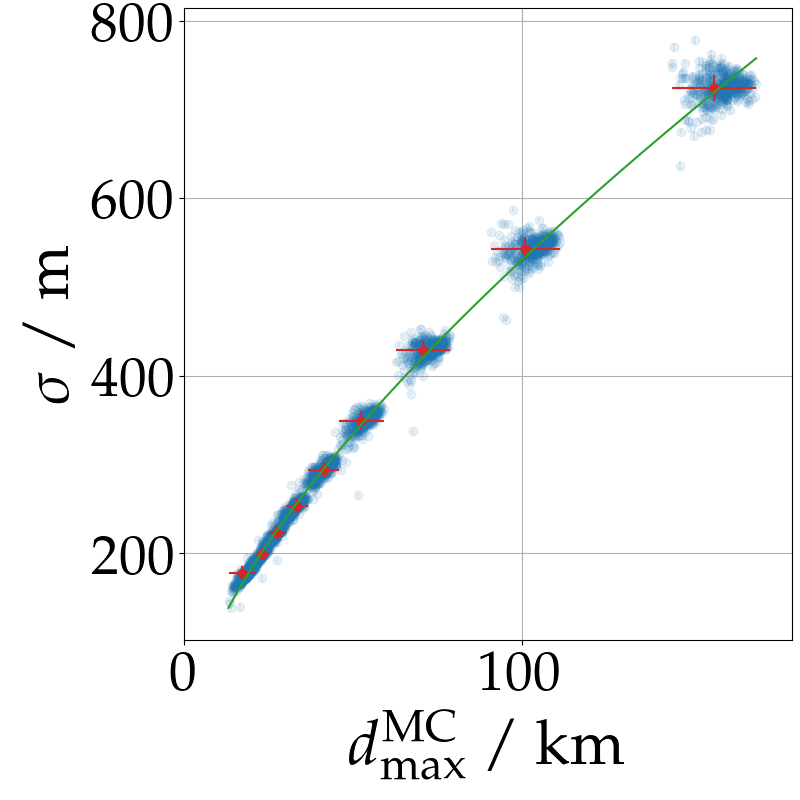}
    \includegraphics[width=0.245\textwidth]{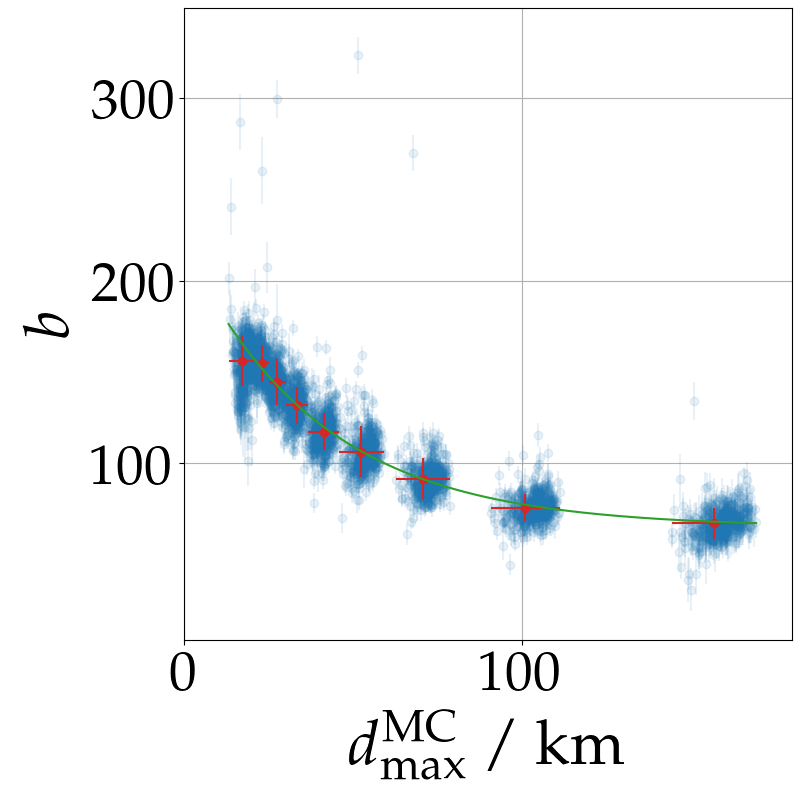}
    \includegraphics[width=0.245\textwidth]{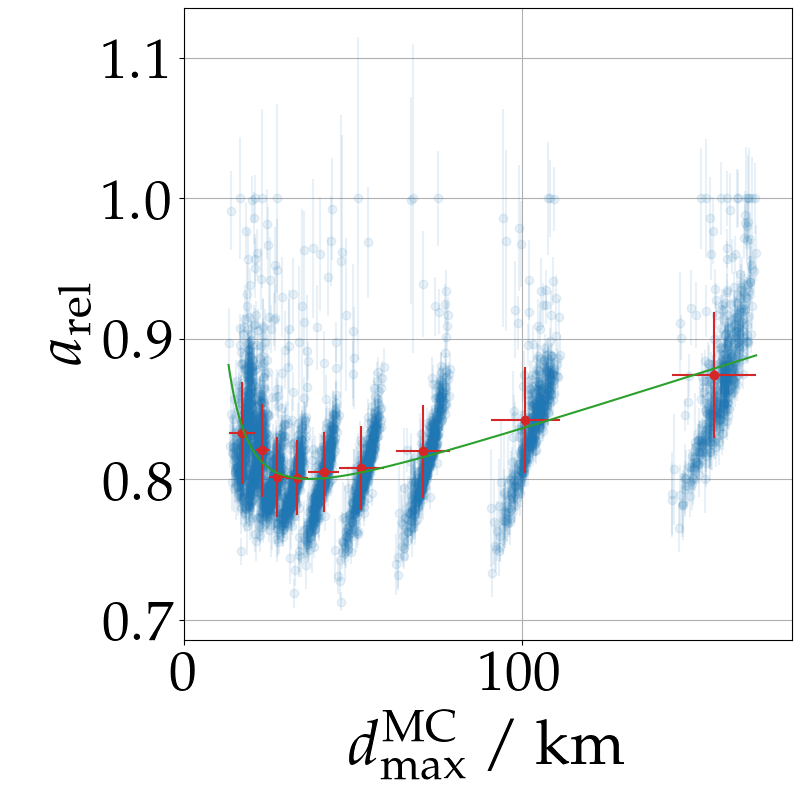}
    \includegraphics[width=0.245\textwidth]{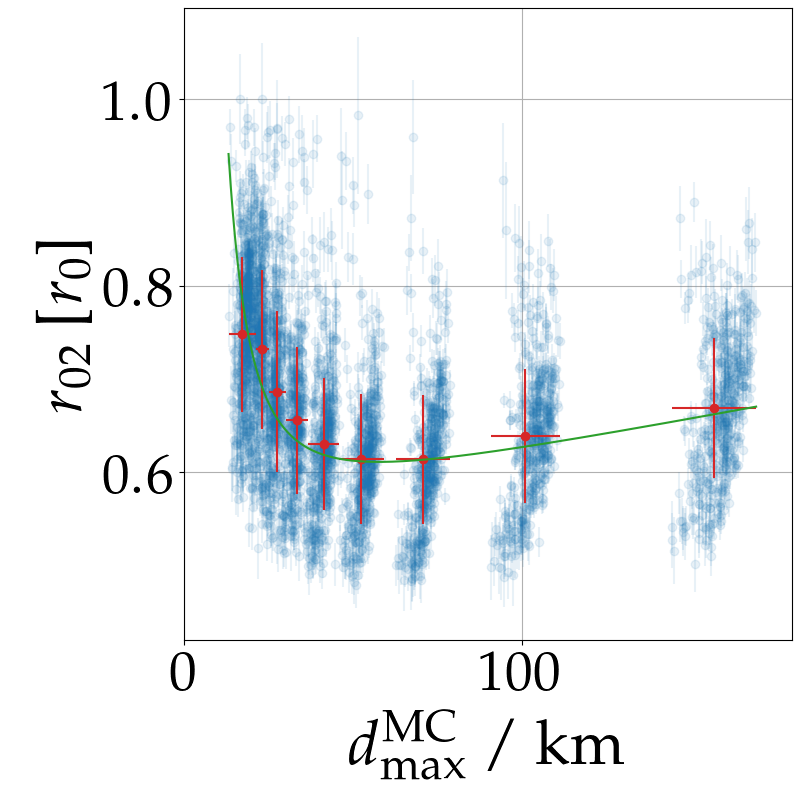}
    \caption{Parameterizations according to eqns.\ (\ref{eqn:sigma}) to (\ref{eqn:r02}) (green lines) compared with the fit values for individual simulations (blue points) as well as their profiles (red points; means and standard deviations).}
    \label{fig:parameterizations}
\end{figure}

For a practical reconstruction algorithm, we need to reduce the number of free parameters in the LDF. We do so by establishing correlations with air-shower characteristics using a library of 4309 air showers initiated by proton and iron nuclei, with energies from \eV{18.4} to \eV{20.2} in bins of $\log_{10}(E / \,\mathrm{eV}) = 0.2$ and zenith angles from 65 to 85$^\circ$ in bins of 2.5$^\circ$ simulated with CoREAS \cite{HuegeCoREAS2012} (pre-release version of CORSIKA V7.7000). We use the interaction models QGSJETII-04 and UrQMD, star-shape antenna grids with 240 antennas, and optimized thinning at a level of $5 \times 10^{-6}$. In the following, we exclude the 240 showers with geomagnetic angles $\leq 20^\circ$. For cross-checks we use an extra 648 showers simulated with three different atmospheres.

As we will see, the primary variable governing the shape of the LDF is $d_\mathrm{max}$, the geometrical distance between the impact point of the shower on the ground (``core'') and the position of the air-shower maximum. It turns out that the position of the maximum of the Gaussian, $r_0$, agrees well with the position calculated from the Cherenkov ring at an angle of $\delta_\mathrm{che}$ as calculated using the refractive index $n$ at the shower maximum:
\begin{equation}
    r_0 = \tan(\delta_\mathrm{che}) \cdot \dmax \, \text{,} \quad \;\; \delta_\mathrm{che} = \cos^{-1}\left(\frac{1}{n(h(\dmax))}\right)\label{eqn:Cherenkov}
\end{equation}
Here, $h$ denotes height above sea level. The agreement between the thus-calculated $\delta_\mathrm{che}$ values and $\delta_\mathrm{Che}(r_0)$ extracted from the fits of individual simulations is demonstrated in Figure \ref{fig:cherenkovangle}. This allows us to simply calculate $r_0$ given the atmospheric model and the position of shower maximum, with sufficient quality as to not deteriorate the reconstruction resolution. Note that $r_0$ and thus $\delta_\mathrm{Che}$ denote the position of the maximum of the Gaussian in eqn.\ (\ref{eqn:fgs}), \emph{not} the position of the maximum of the fluence distribution, as also visible in Figure \ref{fig:LDFs} (right).

We fix parameter $s \equiv 5.0$ to restrict the sigmoid's influence to the inner part of the LDF, and determine the dependence of the remaining parameters on $d_\mathrm{max}$ with an iterative fit procedure. This procedure yields the following parameterizations for the remaining four parameters:

\begin{equation}
    \sigma = \left(0.132 \cdot \left(\frac{\dmax - 5\,\mathrm{km}}{\mathrm{m}}\right) ^ {0.714} +  56.3\right)\,\mathrm{m} \label{eqn:sigma}
\end{equation}

\begin{align}
    p(r) = \left\{ \begin{array}{cc} 
                2 & \hspace{5mm} r \leq r_0 \\
                2 \cdot (r_0 / r) ^ {b / 1000} & \hspace{5mm} r > r_0 
                \end{array} \right. \text{,} \;\; b = 154.9 \cdot \exp\left(-\frac{\dmax}{40.0\,\mathrm{km}}\right ) + 64.9,
\end{align}

\begin{equation}
    a_\mathrm{rel} = 0.757 + \frac{\dmax}{1301.4\,\mathrm{km}} + \frac{19.8\, \mathrm{km}^2}{\dmax^2},
\end{equation}

\begin{equation}
    r_{02} = 0.552 + \frac{\dmax}{1454.2\,\mathrm{km}} + \frac{66.2\, \mathrm{km}^2}{\dmax^2}. \label{eqn:r02}
\end{equation}

The agreement of these parameterizations with values extracted from fitting the LDFs of individual simulations is shown in Figure \ref{fig:parameterizations}. The slanted structures seen in particular for $a_\mathrm{rel}$ and $r_{02}$ stem from variations in depth of shower maximum. We tried a second-order correction for these, which did, however, not improve reconstruction quality. We thus prefer to keep the parameterizations simpler by not including a second-order correction.

The two remaining fit parameters for the LDF in equation (\ref{eqn:fgs}) thus constitute an amplitude $f_0$ and the distance to shower maximum $d_\mathrm{max}$. Also, two angles for the arrival direction (routinely available from a timing fit of radio or particle data) and two core coordinates need to be constrained.

\section{Charge-excess parameterization}

\begin{figure}
    \centering
    \includegraphics[width=0.49\textwidth]{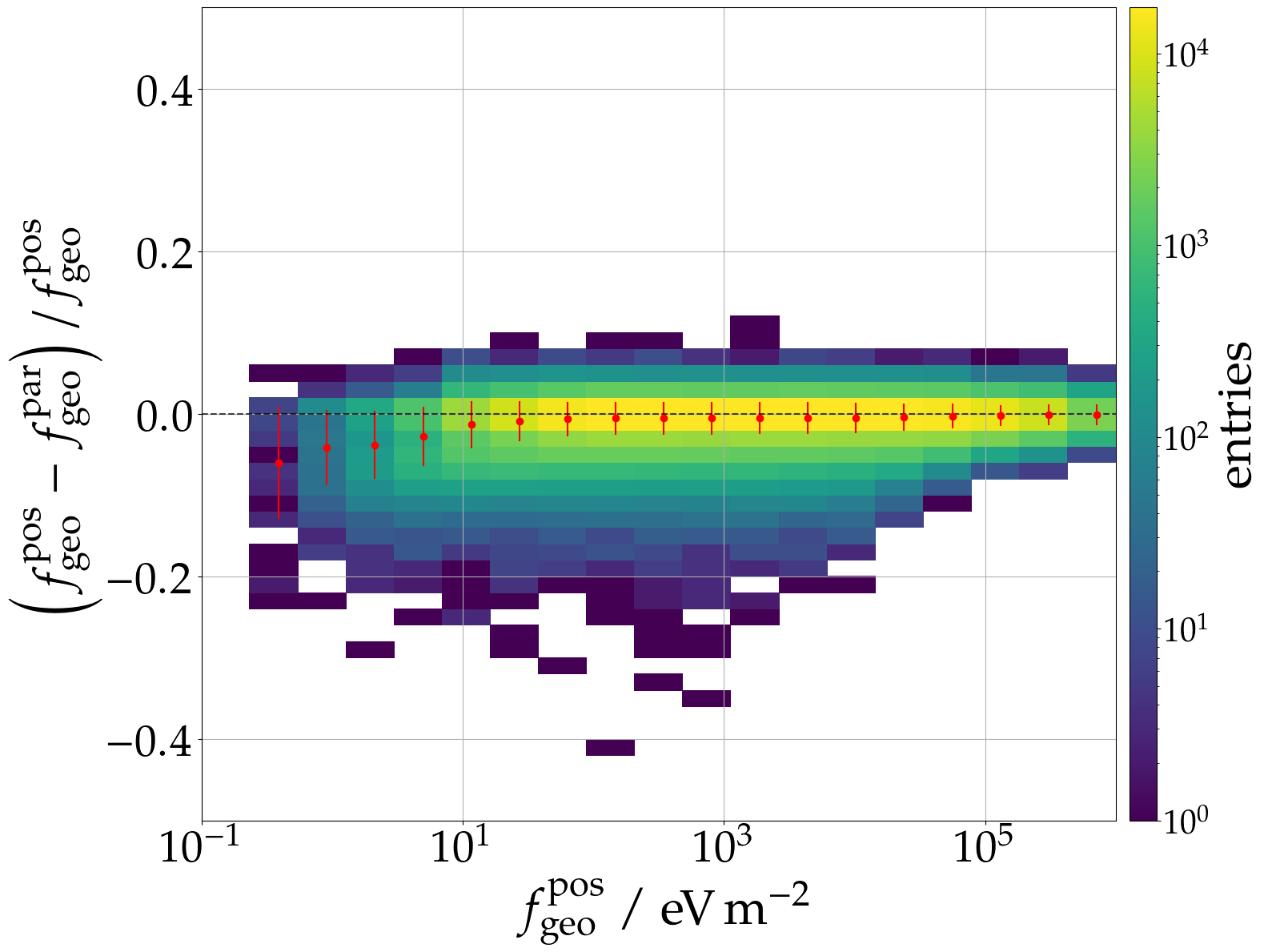}
    \includegraphics[width=0.49\textwidth]{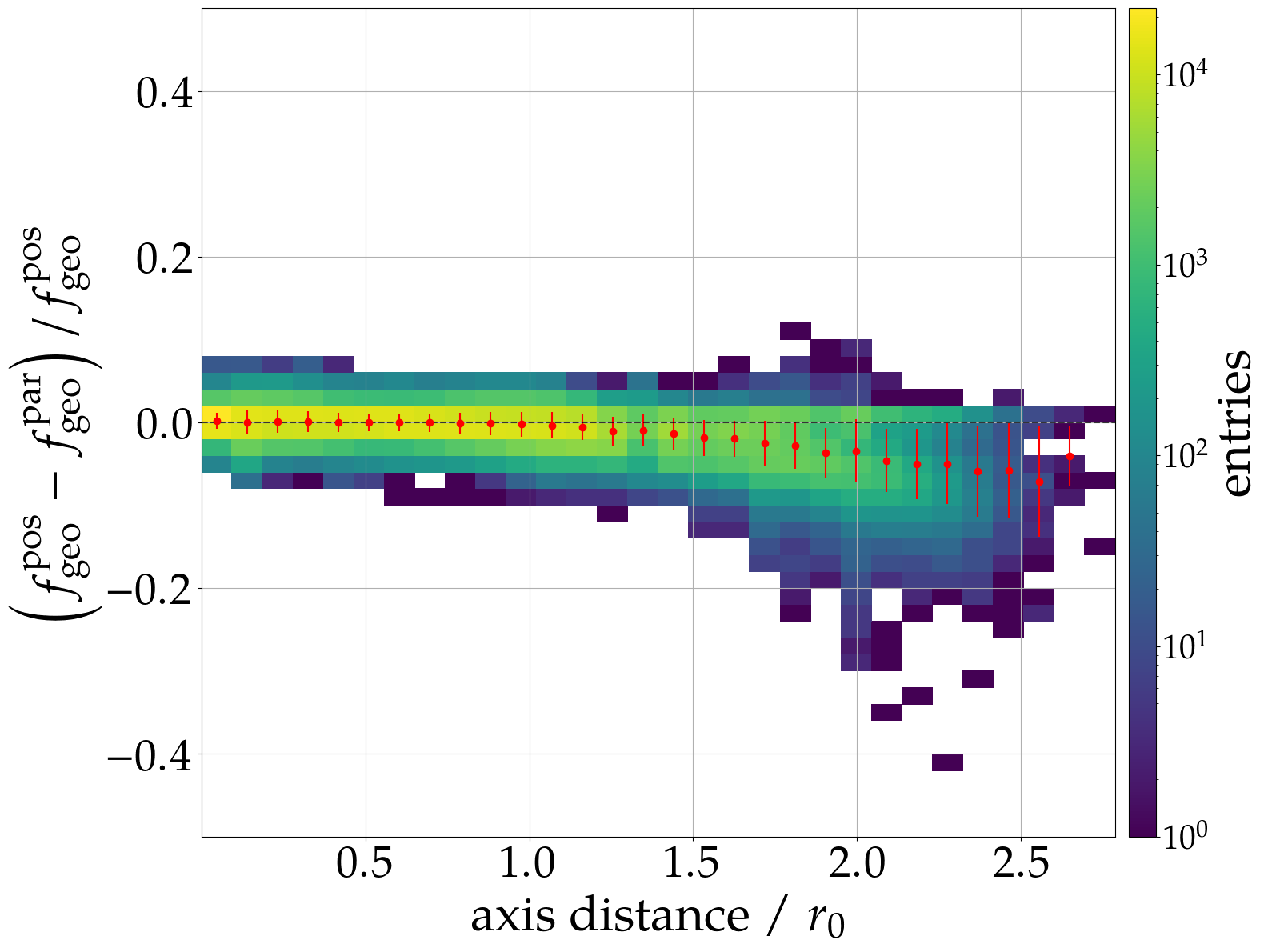}
    \caption{Comparison of the geomagnetic energy fluence $f_\mathrm{geo}^\mathrm{par}$ extracted using eqns.\ (\ref{eqn:aceparam}) and (\ref{eq:geomag}) and $f_\mathrm{geo}^\mathrm{pos}$ as extracted using polarization information per antenna position according to eqn.\ (\ref{eqn:posfraction}). The profile shows means and standard deviations. The bias in the parameterized values is less than 0.2\% and the scatter is within 2\%.}
    \label{fig:chargeexcess}
\end{figure}

Up to this point, we calculated \fgeo by removing the charge-excess fluence for each individual simulated antenna according to the expected polarization at its position using eqn.\ (\ref{eqn:posfraction}). For measured data, this approach will be unreliable because of the presence of noise. We thus parameterize the charge-excess fraction $\ace \equiv \sin^2 \alpha \cdot \frac{\fce}{\fgeo}$, where $\alpha$ is the ``geomagnetic angle'', as
%
\begin{equation}
    \ace = \left[0.348 - \frac{\dmax}{850.9\,\mathrm{km}} \right]  \cdot \frac{r}{\dmax} \cdot \exp\left[\frac{r}{622.3\,\mathrm{m}}\right] \cdot \left[ \left( \frac{\rhomax}{0.428 \, \mathrm{kg}\,\mathrm{m}^{-3}} \right) ^ {3.32} - 0.0057 \right]. \label{eqn:aceparam}
\end{equation}
%
To derive this parameterization, simulated pulses significantly affected by thinning have been excluded. $\rho_\mathrm{max}$ denotes the atmospheric density at the position of the shower maximum, which can be calculated directly from $d_\mathrm{max}$ and the zenith angle for any given atmospheric model.

This parameterization has been improved with respect to our earlier one \cite{HuegeIcrc2019}, in particular it now holds up to 85$^\circ$ zenith angle (previously 80$^\circ$). Using this parameterization, the geomagnetic energy fluence can thus be extracted from the fluence in the $\bf{v} \times \bf{B}$ polarization as
\begin{equation}
    f_{\mathrm{geo}}^{\mathrm{par}} = \frac{f_{\textbf{v}\times\textbf{B}}}{\left(1 + \frac{\cos(\phi)}{|\sin(\alpha)|} \cdot \sqrt{a_\mathrm{ce}(r, d_{\mathrm{max}}, \rho_{\mathrm{max}})}\right)^2}, \label{eq:geomag}
\end{equation}
the agreement of which with values extracted directly according to eqn.\ (\ref{eqn:posfraction}) is shown in Figure \ref{fig:chargeexcess}. 

\section{Energy reconstruction on a 1.5 km grid}

\begin{figure}
    \centering
    \includegraphics[width=0.99\textwidth]{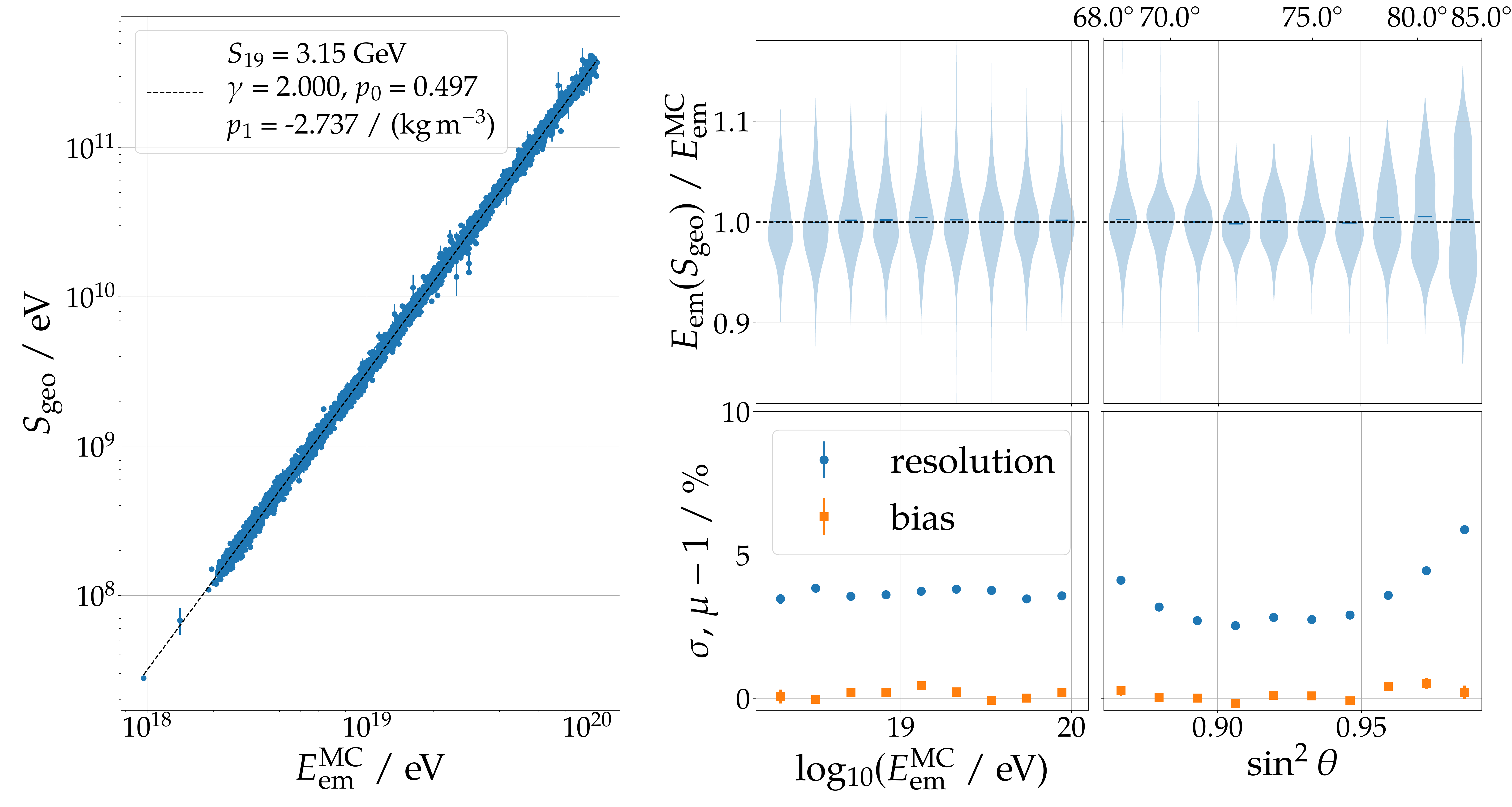}
    \caption{Left: Correlation of the corrected geomagnetic radiation energy $S_\mathrm{geo}$ as reconstructed from individual simulations on a 1.5 km grid compared with the corresponding Monte-Carlo true electromagnetic energy. The legend shows the fit parameters when fitting the data with eqn.\ (\ref{eq:Scorrelation}). Middle: Comparison of electromagnetic energy $E_\mathrm{em}$ reconstructed using the quoted fit parameters and eqn.\ (\ref{eq:Scorrelation}) compared to Monte Carlo truth as a function of energy. In the lower panel, bias and resolution are shown. Right: Same as middle panel but as a function of zenith angle.}
    \label{fig:rec}
\end{figure}

We now have all ingredients to try and reconstruct individual air showers. We use a second library of 6060 CoREAS simulations for proton, helium, nitrogen and iron primaries with energies between \eV{18.4} and \eV{20.1} randomly distributed in $\log_{10}(E /\,\mathrm{eV})$ and isotropic arrival directions with zenith angles between 68.0 and 85.0$^\circ$, which are no longer on a star-shape grid but on a 1.5\,km hexagonal grid of antennas. We only use simulations for which at least 5 positions have been simulated. Arrival directions with $\alpha \leq 20^\circ$ are not included in this set. These simulations have been performed with an optimized thinning at the $10^{-6}$ level. We do not add noise to our simulations nor conduct any detector simulation as to judge the intrinsic achievable resolution of the method. For the direction we use the true Monte Carlo value; for the core position we use the Monte Carlo values as starting values in the fit procedure. 

The first step is to determine the ``geomagnetic radiation energy'' of a given shower via area-integration of the energy fluence. To include this explicitly as parameter \Egeo, we rewrite our LDF such that we normalize out the integral over the fit function (numerically),
\begin{equation}
    \fgeo(r) = \Egeo \frac{\fgs(r)}{2 \pi \int_0^{5r_0} \fgs(r) r \, \mathrm{d}r },
\end{equation}
%
where $f_0$ was now set to unity. The effective fit parameters are two core coordinates in the shower plane, $d_\mathrm{max}$ and \Egeo. The fit succeeds with high quality for 6002 of the 6060 simulated showers. The fitted geomagnetic radiation energy is then converted to ``corrected geomagnetic radiation energy'' by compensating for the geomagnetic angle, magnetic field strength and air density at shower maximum (following the logic established in \cite{GlaserJCAP2016}):
\begin{equation}
        S_{\mathrm{geo}} = \frac{E_{\mathrm{geo}}}{\sin^2(\alpha)} \cdot
        \frac{1}{\left(1 - p_0 + p_0 \cdot \exp\left[p_1 \cdot (\rho_\mathrm{max} - \langle \rho \rangle)\right]\right)^2} \label{eq:corrS}
\end{equation}
Note that we normalize to a mean air density of $\langle \rho \rangle = 0.3\,$kg/m$^3$ which is an adequate average value for inclined air showers. (For vertical showers, a value of $\sim 0.65$\,kg/m$^3$ was used so far \cite{HuegeIcrc2019,GlaserJCAP2016}.) The choice of this value affects the interpretation and value of $S_{19}$ defined below. We note that also the magnitude of the magnetic field can be normalized out, using a power-law with $B^{1.8}$ \cite{GlaserJCAP2016}.

As a final step, we correlate the corrected geomagnetic radiation energy with the electromagnetic energy $E_{\mathrm{em}}$ of the shower, extracted from the CORSIKA longitudinal file by summing over the complete energy deposit columns of electrons and positrons. Clipping of showers is irrelevant for these inclined geometries. The correlation follows:
\begin{equation}
S_\mathrm{geo} = S_{19} \cdot \left(\frac{E_{\mathrm{em}}}{10 \, \text{EeV}}\right)^\gamma \label{eq:Scorrelation}
\end{equation}
In Figure \ref{fig:rec} we illustrate the quality of the achieved reconstruction for our simulation library. The left diagram shows a scatter plot of corrected geomagnetic radiation energy versus Monte Carlo true electromagnetic energy. The middle diagram illustrates the agreement between reconstructed and Monte-Carlo true electromagnetic energy as a function of Monte Carlo electromagnetic energy in terms of distributions, resolution $\sigma$ and bias $\mu$. The resolution is consistently better than 5\% and the bias is negligible. The right panel illustrates the same performance as a function of air-shower zenith angle. A slight but acceptable loss of resolution is visible towards the lowest and highest zenith angles. We stress that in an actual experiment, performance will degrade due to the presence of noise, an imperfect knowledge of the response of individual antennas, and the degradation of the available information on the arrival direction and core position. A study for realistic conditions as adequate for the AugerPrime Radio Detector is presented in reference \cite{SchlueterIcrc2021}. The values we quote here demonstrate that the \emph{intrinsic} resolution in electromagnetic energy achievable by our reconstruction algorithm is better than 5\% at all energies.

\section{Conclusions}

Using CoREAS simulations, we have developed a reconstruction algorithm for radio measurements of inclined air showers. It relies on symmetrizing the signal distribution in terms of the geomagnetic energy fluence \fgeo by accounting for refractive core displacement, geometrical early-late corrections, and asymmetries introduced by the superposition of geomagnetic and charge-excess emission. The latter is removed using a parameterization of the charge-excess fraction. The symmetrized, geomagnetic, energy fluence is then fit with a one-dimensional (rotationally symmetric) lateral distribution function which is the sum of a Gaussian and a sigmoid component. We have successfully parameterized most parameters of the LDF as a function of the geometrical distance of the shower maximum from the core. The remaining free fit parameters are the geomagnetic radiation energy and the geometrical distance of the shower maximum, along with two core coordinates. Furthermore, the arrival direction and atmospheric model must be known. The geomagnetic radiation energy arising from the fit is then corrected for density and geometry effects, yielding an energy estimator that scales quadratically with the electromagnetic energy of the air shower. For air showers initiated by protons, helium, nitrogen and iron nuclei with zenith angles between 68 and 85$^\circ$ sampled on a 1.5\,km antenna grid, the electromagnetic energy can be reconstructed with an intrinsic resolution of 5\% at all investigated energies.

Our approach is directly applicable to the AugerPrime Radio Detector and should lend itself well also to other experiments focused at radio detection of inclined air showers such as GRAND when adapting it to the changed environmental conditions and observing frequency band.

\acknowledgments
We would very much like to thank Alan Coleman for the idea to try a Gauss-sigmoid LDF and for many fruitful discussions. Felix Schlüter is supported by the Helmholtz International Research School for Astroparticle Physics and Enabling Technologies (HIRSAP) (grant number HIRS-0009). Simulations for this work were performed on supercomputers BwUniCluster 2.0 and ForHLR II at KIT funded by the Ministry of Science, Research and the Arts Baden-Württemberg and the Federal Ministry of Education and Research. The authors acknowledge support by the state of Baden-Württemberg through bwHPC.
\end{document}